# The Computation of Cyclic Peptide with Prolin-Prolin Bond as Fusion Inhibitor of DENV Envelope Protein through Molecular Docking and Molecular Dynamics Simulation

Arli Aditya Parikesit[1], Hilyatuz zahroh[1], Andreas S Nugroho[1], Amalia Hapsari[1], Usman Sumo Friend Tambunan[1]*

[1]Department of Chemistry, Faculty of Mathematics and Science, University of Indonesia

* Corresponding author: usman@ui.ac.id

A disease that caused by dengue virus (DENV) has become the major health problem of the world. Nowadays, no effective treatment is available to overcome the disease due to the level of dengue virus pathogeneses. A novel treatment method such as antiviral drug is highly necessary for coping with the dengue disease. Envelope protein is one of the non-structural proteins of DENV, which engaged in the viral fusion process. It penetrates into the host cell to transfer its genetic material into the targeted cell followed by replication and establishment of new virus. Thus, the envelope protein can be utilized as the antiviral inhibitor target. The fusion process is mediated by the conformational change in the protein structure from dimer to trimer state. The previous research showed the existing cavity on the dimer structure of the envelope protein. The existing ligand could get into cavity of the envelope protein, stabilize the dimer structure or hamper the transition of dimer protein into trimer. In this fashion, the fusion process can be prevented. The aim of this research is designing the cyclic peptide with prolin-prolin bond as fusion inhibitor of DENV envelope protein through molecular docking and molecular dynamics simulation. The screening of 3,883 cyclic peptides, each of them connected by prolin-prolin bond, through molecular docking resulted in five best ligands. The pharmacological and toxicity character of these five ligands were analised in silico. The result showed that PYRRP was the best ligand. PAWRP was also chosen as the best ligand because it showed good affinity with protein cavity. Stability of ligand-protein complex was analyzed by molecular dynamics simulation. The result showed that PYRRP ligand was able to support the stability of DENV envelope protein dimer structure at 310 K and 312 K. While PAWRP ligand actively formed complex with the DENV envelope protein at 310 K compared to 312 K. Thus the PYRRP ligand has a potential to be developed as DENV fusion inhibitor.

Keywords: dengue, envelope protein, fusion process, cavity, cyclic peptide, molecular docking, molecular dynamics

## 1. Introduction

Disease that caused by dengue virus infection has become a global health problem, especially in tropical and subtropical regions such as Asia, Africa, and America. This infection has become endemic in more than 100 countries, including Indonesia. The World Health Organization estimates that there have been 50-100 million cases of dengue virus infection each year and as many as 2.5 billion people or 40% of the world's population are at risk of suffering this virus infection [1].

DENV, which is a positive single-chain RNA virus, belongs to the genus flavivirus and family Flaviviridae. DENV has four serotypes, namely DENV-1, DENV-2, DENV-3 and DENV-4. Those serotypes had the same morphology and genome but show different antigens, thus a person can be infected more than once as there is no complete antibodies cross protection [2].

Dengue virus RNA genome is a single strand Open Reading Frame (ORF) that encodes a polyprotein consisting of 3,391 amino acid residues forming three structural proteins, namely: C (capsid), PRM (pre-membrane), and E (envelope), and seven non-structural (NS) protein NS1, NS2A, NS2B, NS3, NS4A, NS4B, and NS5 [3].

DENV life cycle begins with the attachment of virions to the host cell receptors and then enters the cell by endocytosis. The decrease in the pH of the endosome triggers the trimerization of Virus E protein that is irreversible and leads to the fusion process between the viral and host cell membranes. Through this process, the viral genome can be transferred into the cell.

The fusion process is catalyzed by a change in condition of low pH in the endosome (~5.8-6.0), causing a conformational change in the E dimer protein structure. Low pH condition in the endosome causes protonation of histidine residues that conserved in specific E protein. It triggers dimer protein dissociation of E structure. E protein composed of two identical chains. Each chain consists of three domains, they are: Domain I, N-terminal portion located at the tip; Domain II, the area of the fusion, and Domain III, is an area of the receptor that bind to the host cells (Fig. 1).

The fusion process occurs at the end of the fusion peptide loop of domain II. It occurs due to the movement around the hinge region between domains I-II border. This exposure led to the entry of the fusion peptide into the host cell membrane (Fig. 2). At this transition stage, E protein of the virus was linked with the host cell membrane. Both membranes are close together while the fusion of E protein triggers the formation of trimer [4]. Previous studies related to the inhibition of DENV E protein by Modis et al., [5] found the existence of a binding region (binding pocket) on the structure of the DENV-2 E protein. This pocket is occupied by a detergent molecule known as n-octyl-β-D-glucoside (β-OG). Hence this pocket is called as β-OG pocket and located in the hinge region between the envelope protein domains I-II. Some related researches focused on finding inhibitors of the fusion process which can bind to β-OG binding pocket. The presence of the ligand occupying this site is expected to inhibit or prevent the process of viral-host fusion and trimerization of E protein. Transition of trimer formation is an important step in the entry of the dengue virus genome into the host cell [6,7,8].

Other related studies have also been conducted. Yennamali et al., [9] have found the cavity site between domains I-III on the DENV-2 E protein. The site is only found in the structure of dimeric E protein. The presence of a ligand that occupies this cavity stabilizes the dimer structure or inhibits E protein trimerization, so that the fusion process can be disturbed. This research focused to find a compound that can occupy this cavity through virtual screening methods.

In our previous researches, we have designed disulfide cyclic peptide to target NS2B-NS3 protease, RNA-dependent RNA polymerase and methyltransferase of NS5 protein of DENV [10, 11, 12, 13]. In this research, we designed cyclic peptides containing proline - proline bond and tested their affinities to target cavity of E protein via the molecular docking and molecular dynamics simulations. The selection of proline residues to form cyclic peptide was based on the ability of the secondary amine group of proline in reducing the rotational energy of proline - proline mediated-cyclization [14].

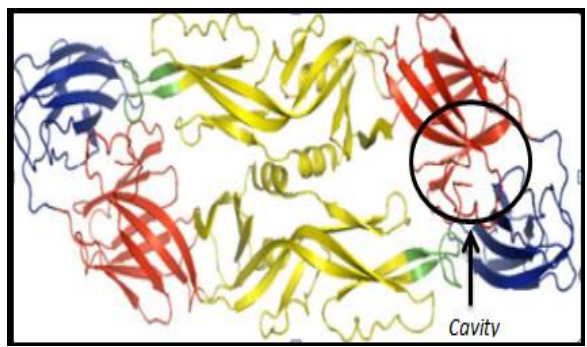

*Fig 1.* DENV envelope protein dimer structure and the location of the cavity; domain I (red), domain II (yellow,) fusion peptide (green), and domain III (blue)

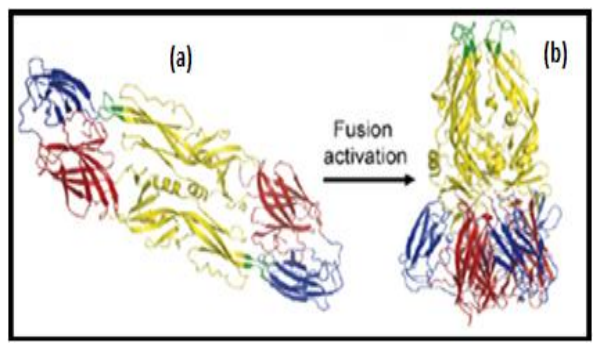

*Fig 2.* E protein trimerization process (fusion activation); (a) dimer structure, (b) trimer structure

## 2. Methodology

**Preparation of the envelope protein (E) DENV**

E Protein with 1OAN_A code was downloaded from the PDB and opened in the Molecular Operating Environment (MOE) 2008.10. E protein structure was adjusted and optimized by protonate 3D option in MOE [10]. The addition of hydrogen atoms was done by selecting the partial charge option. Energy minimization process was conducted by using the AMBER99 force field, gas phase solvation. The value of RMS gradient was set to 0.05 kcal / mol Å [13].

**Design and preparation of cyclic peptide ligands**

The determination of cyclic peptide ligands as inhibitors was performed by analyzing the amino acid residues of the E protein cavity. The cyclic pentapeptide was formed with proline residues at its terminal. The combination of 7 polar amino acids (glycine, serine, threonine, cysteine, tyrosine, asparagine, and glutamine), 8 non polar amino acids (alanine, valine, leucine, isoleucine, phenylalanine, tryptophan, methionine, and proline), and 5 charged polar amino acids (lysine, arginine, histidine, aspartic acid, and glutamic acid) generated 3,883 cyclic pentapeptide molecules.

The process of designing 3D structure of the ligand was done by using ChemSketch ACDLabs. Ligand optimization and minimization was done with MOE 2008.10. Optimization process was utilized by selecting the wash option and the partial charge option for all designed ligands. Energy minimization process was applied by using MMFF94x force field, gas phase solvation, and the value of RMS gradient 0.001 kcal / mol Å [13].

**Molecular Docking**

The docking between cyclic peptides as the ligands and E protein was done by selecting simulation_dock option in MOE. Triangle matcher was selected as placement method. The utilized scoring function is London dG. The next appearance of the 100 best pose was refined based on force field parameter. The displayed result of overall docking process is the best pose.

**Predicted pharmacological properties of the Ligands**

Prediction of the pharmacological properties of the cyclic peptides was applied to the ligands that came out as the best results of molecular docking. The designated pharmacological properties are Lipinski's Rules of five, oral bioavailability, druglikeness, and drug score. Predictions were performed by using Osiris Property Explorer and FAF-drugs2 software.

**Toxicity studies of the cyclic peptide ligands**

Toxicity studies were also conducted on the best ligands from molecular docking results. Predictive nature of the ligand carcinogenic and mutagenic properties was performed using software Toxtree v-2.5.0 based on Benigni and Bossa rules. The performed predictions are mutagenic, tumorigenic, irritant and reproductive effective characters of the ligands by using Osiris Property Explorer software. Analysis of the ADME-Tox (Absorption, Distribution, Metabolism, Excretion and Toxicity) of the ligand was performed using ACD software I-labs/Percepta.

**Molecular Dynamics Simulation**

Optimization and energy minimization of protein-ligand complexes were required before performing molecular dynamics simulation. Optimization was done by selecting the partial charge option. Energy minimization was applied by using the AMBER99 force field and born solvation. The RMS gradient value was set to 0.05 kcal/molÅ. Molecular dynamics simulation was performed using MOE by selecting simulation_dynamic option. The utilized parameters were NVT ensemble and the NPA algorithm. The applied force field was AMBER99.

**3. Results and Discussion**

**Cavity analysis of E DENV protein**

The cavity of E DENV protein was composed of 25 amino acid residues: residues 1, 143-149, 156, 158, 178 and 295 of domain I, and residues 324, 333, 355-357, 359-366 of domain III [9]

(Fig. 3). Based on the visualization cavity with MOE, polar; uncharged polar; and non-polar amino acid residues selected as constituents of cyclic peptides (Fig. 4).

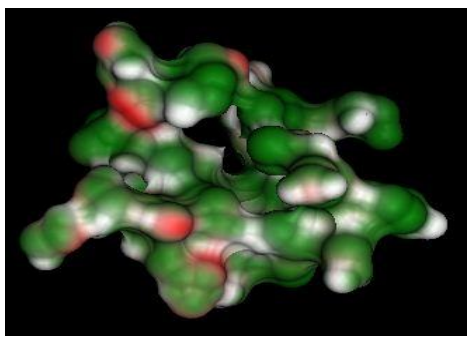

*Fig 3.* Cavity visualization of E protein DENV

**Ligand screening through molecular docking process**

*Table 1.* Five best cyclic peptide ligands according to molecular docking result

| Ligand | $\Delta G_{binding}$ (Kkal/mol) | $pK_i$ |
|---|---|---|
| PYRRP | -24.08 | 17.54 |
| PAWRP | -20.29 | 14.78 |
| PCWRP | -19.96 | 14.54 |
| PFWRP | -18.94 | 13.79 |
| PWPRP | -18.39 | 13.39 |
| R1 Yennamalli* | -14.35 | 10.45 |
| CLREC* | -12.36 | 9.00 |
| A4 Kampmann* | -15.52 | 11.31 |
| A5 Kampmann* | -14.69 | 10.69 |
| C6 Wang* | -13.86 | 10.10 |
| NITD448 Poh* | -11.567 | 8.423 |

*: Standard

Screening of 3,883 cyclic peptide ligands resulted in 5 best ligands based on the value of Gibbs free energy ($\Delta G_{binding}$) and the value of $pK_i$ (Table 1). Based on the thermodynamic equation, there is a relationship between the value of the binding free energy ($\Delta G_{binding}$) and the inhibition constant ($K_i$) [15]:

$$\Delta G^o = -RT \ln K_A \qquad K_A = K_i^{-1} = \frac{[EI]}{[E][I]}$$

$K_A$ = Biological Activity constant  [14]
$K_i$ = Inhibition constant

Stability and strong interactions in protein-ligand complexes can be seen from the negative value of the Gibbs free energy of binding ($\Delta G_{binding}$) that displayed on the final docking result. Negative values indicate that the reaction of protein-ligand complex formation takes place spontaneously. $K_i$ values are shown as the mean value of the $pK_i$, minus the value of the logarithm of $K_i$. The greater the value of $pK_i$, the more stable the complex of protein-ligand.

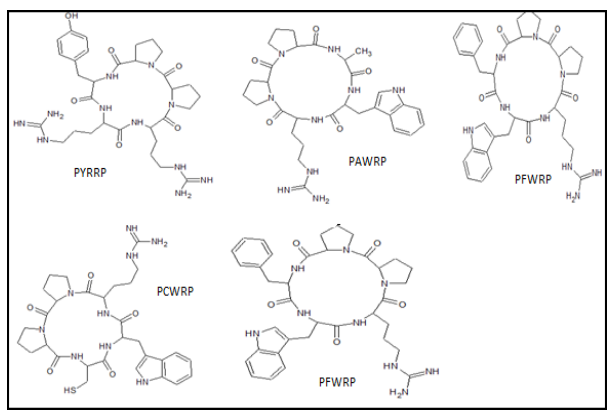

*Fig 4.* 2D images of the best cyclic peptide ligand screening results by molecular docking

According to the value of ΔG$_{binding}$ and pKi, the PAWRP and PYRRP ligands showed better stability and good interaction with the protein than 6 standard ligands (Table 1). Stability and good interaction of the complex formation can be observed from the interaction of the hydrogen bonds.

**The prediction of the ligand's pharmacological properties**

There are four parameters of the Lipinski's rule of Five that is needed to fulfil, they are: the molecular weight of the molecule does not exceed 500 Da, logP value is not higher than 5, the number of H-donor (n OHNH) is less than 5, and the number of H-acceptor (n ON) is less than 10 [16]. Nonetheless, Lipinski's rule does not always serve as absolute reference for drug design. Generally, Lipinski's Rule of Five is applied to predict druglikeness of a molecule to see if it is orally active drug.

Five cyclic peptide ligands (PYRRP, PAWRP, PCWRP, PFWRP, and PWPRP) did not meet the criteria of Lipinski's Rule. The ligands are cyclic pentapeptide whose size and molecular weight lie outside the range of Lipinski's rule (Table 2). Partition coefficient (logP) value is defined as the ratio of the concentration of a molecule in octanol and water. LogP value is associated with the hydrophobicity of a drug molecule. As the value of logP increases, the hydrophobicity character of the compound also increases. According to Lipinski's Rule, five cyclic peptide ligands have logP values less than 5. The nature of the ligand is influenced by the hydrophobicity of side chain groups of amino acid residues that make up the ligand.

High oral bioavailability is an important consideration in developing bioactive molecules as therapeutic agents. The prediction of oral bioavailability is based on the parameters of Veber's rules, which includes: the number of rotatable bonds (n rotb) is 10 or fewer, topological polar surface area (TPSA) is less than 140 Å and the number of H-bonds (H-donor and H-acceptor) is less than 12. TPSA is defined as the surface area of all polar atoms of drug compound that is affected by oxygen (O), nitrogen (N), and hydrogen (H) atoms. TPSA values tend to be used as a matrix for optimization of drug traffic that penetrates the cell membrane [17].

The prediction of oral bioavailability showed that all five cyclic peptide ligands have low oral bioavailability compared with standard ligands (Table 2). The low oral bioavailability is due to the domination of polar amino acids that compose the ligands, consequently TPSA value of the ligands is greater than the upper limit set by Veber's rule.

The drug likeness of PYRRP, PAWRP, PFWRP, PCWRP, and PWPRP ligands is greater than the standards. The drug score is a combination of druglikeness, logP, solubility, molecular weight, and toxicity risk that is used to assess the overall potential qualification of the compound as a drug. PYRRP, PAWRP, PCWRP, and PWPRP ligands' drug score is greater than the standards.

**Toxicity studies of the ligand**

Toxtree prediction analysis showed that the ligands (PYRRP, PAWRP, PCWRP, PFWRP, and PWPRP) do not possess mutagenic and carcinogenic properties. It was computed based on QSAR (Quantitative Structure-Activity-Relationship) approach. The five ligands also did not have a tendency to be carcinogenic in both genotoxic and non-genotoxic levels (Table 3).

The result of toxicity prediction using Osiris Property Explorer was shown by the color parameter; the green color indicates a low level of propensity of the parameter, while the red color indicates a high propensity of the parameter. Prediction results indicated that the PYRRP, PAWRP, and PWPRP ligands have a low tendency towards mutagenic, tumorgenic, irritant, or reproductive effective properties (Table 4). As for the PFWRP and PCWRP ligands, they have a high level of propensity to mutagenic properties. Both ligands possess aromatic ring of phenylalanine (F) and tryptophan (W), which is toxic to the body, on their amino acid residues.

The result of ADME-Tox properties analysis showed that the five cyclic peptide ligands (PYRRP, PAWRP, PCWRP, PFWRP, and PWPRP) have side effect on the circulatory system and on several organs including liver, kidney and intestine. This may be due to the aromatic ring in the side chain groups of amino acid residues phenylalanine (F) and tryptophan (W) that make up the ligand. The large size of the ligands might also result in the blockage of arteries. Even so, the five designed cyclic peptide ligands have more toxic probabilities than standard ligands (Table 5). ADME-Tox analysis result also showed that oral bioavailabilty character of the tested ligands is low (<30%) (Table 5). This is consistent with the predictions of pharmacological properties based on Veber's Rules [17]. Low oral bioavailabilty requires the drug delivery process through injection.

*Table 2.* The prediction of pharmacogical properties of cyclic peptide ligands

| Ligand | Molecular Weight (Da) | LogP | n ON | n OHNH | tPSA (Å) | n rotb | Drug likeness | Drug score |
|---|---|---|---|---|---|---|---|---|
| PYRRP | 669.77 | -1.07 | 17 | 12 | 271.95 | 12 | 9.44 | 0.55 |
| PAWRP | 607,70 | 0.24 | 14 | 8 | 205.61 | 7 | 8.67 | 0.58 |
| PFWRP | 683.80 | 1.83 | 14 | 8 | 205.61 | 9 | 9.15 | 0.29 |
| PCWRP | 639.77 | 0.16 | 14 | 8 | 244.41 | 8 | 4.06 | 0.32 |
| PWPRP | 633.74 | -0.24 | 14 | 7 | 196.82 | 7 | 9.09 | 0.56 |
| R1 Yennamalli* | 414.84 | 5.84 | 6 | 1 | 72.95 | 4 | 0.98 | 0.26 |
| A4 Kampman* | 402.69 | 4.12 | 6 | 2 | 100.08 | 4 | 2.12 | 0.20 |
| A5 Kampman* | 469.39 | 7.57 | 5 | 1 | 87.64 | 6 | 1.80 | 0.50 |
| C6 Wang* | 428.94 | 5.95 | 4 | 1 | 78.94 | 5 | 0.29 | 0.08 |
| NITD448 Poh* | 653.49 | 7.30 | 6 | 2 | 122.27 | 12 | -4.44 | 0.05 |

Note: *= standard, n ON = the number of H-acceptor, n OHNH = the number of H-donor, n rotb = the number of rotatable bond.

*Table 3.* Toxicity prediction using Toxtree

| Ligand | Negative for genotoxic carcinogenicity | Negative for nongenotoxic carcinogenicity | Potential *S. typhimurium* TA100 mutagen based on QSAR | Potential carcinogen based on QSAR |
|---|---|---|---|---|
| PYRRP | Yes | Yes | No | No |
| PAWRP | Yes | Yes | No | No |
| PFWRP | Yes | Yes | No | No |
| PCWRP | Yes | Yes | No | No |
| PWPRP | Yes | Yes | No | No |
| R1 Yennamalli* | Yes | No | No | No |
| C6 Wang* | Yes | No | No | No |
| A4 Kampmann* | Yes | Yes | No | No |
| A5 Kampmann* | No | No | No | No |
| NITD448 Poh* | No | Yes | No | No |

Note: *: standard

*Table 4.* Toxicity prediction analysis of the cyclic peptide ligands using Osiris Property Explorer

| Ligand | Mutagenic | Tumorgenic | Irritant | Reproductive effective |
|---|---|---|---|---|
| PYRRP | Green | Green | Green | Green |
| PAWRP | Green | Green | Green | Green |
| PFWRP | Red | Green | Green | Green |
| PCWRP | Red | Green | Green | Green |
| PWPRP | Green | Green | Green | Green |
| R1 Yennamalli* | Green | Green | Green | Green |
| C6 Wang* | Green | Yellow | Green | Green |
| A4 Kampmann* | Green | Green | Yellow | Green |

| A5 Kampmann* | Yellow | Red | Green | Green |
| NITD448 Poh* | Green | Green | Red | Green |

Note:  *: standard

*Table 5.* ADME-Tox properties prediction of the cyclic peptide ligands using ACD I-Lab

| Parameter | PYRRP | PAWRP | PFWRP | PCWRP | PWPRP |
|---|---|---|---|---|---|
| Oral Bioavailibility | Less than 30% | Less than 30% | Less than 30% | Less than 30% | Less than 30% |
| Health Effects: | | | | | |
| Blood | 1.00 | 1.00 | 1.00 | 1.00 | 0.96 |
| Cardiovascular | 0.36 | 0.97 | 0.63 | 0.98 | 1.00 |
| Gastrointestinal | 0.90 | 0.96 | 1.00 | 0.98 | 0.97 |
| Kidney | 0.99 | 1.00 | 1.00 | 1.00 | 1.00 |
| Liver | 0.99 | 1.00 | 1.00 | 0.99 | 0.99 |
| Lungs | 0.89 | 0.84 | 0.96 | 0.96 | 0.89 |
| Probability of Toxicity | No (94% probability non-toxic) | Yes (80% probability harmful) | Yes (78% probability harmful) | Yes (85% probability hazard and harmful) | Yes (80% probability harmful) |
| PGP Inhibitor | No | No | No | No | No |
| CNS active | No | No | No | No | No |
| Active Transport : | | | | | |
| Pep T1 | Not Transported | Not Transported | Not Transported | Not Transported | Not Transported |
| ASBT | Not Transported | Not Transported | Not Transported | Not Transported | Not Transported |

Based on the molecular docking, pharmacological and toxicological properties prediction, the PYRRP and PAWRP ligands serve as the best ligand to target DENV envelope protein. This ligands were then proceeded to for molecular dynamics simulation.

**The results of molecular dynamics simulations**

Molecular dynamics simulation was used to analyze the conformational changes in the protein-ligand complex due to the influence of implicit solvent into the system. Therefore, the protein and ligand flexibility was set to born solvation [18].

There are three stages in the molecular dynamics simulations; the initialization, equilibration, and production stage. In the initialization phase, the initial state of the system was determined, including atomic coordinate, velocity and potential energy of the system.

Initialization phase was done at a temperature of 300 K for 100 ps. The simulation was run until the system reached equilibration point that was indicated by the decline in the potential energy of system in order to achieve stability [19]. Production stage of the simulation produced trajectory. The obtained trajectory is a coordinate that was formed by taking the data changes in the state of each atom versus time due to the influence of temperature.

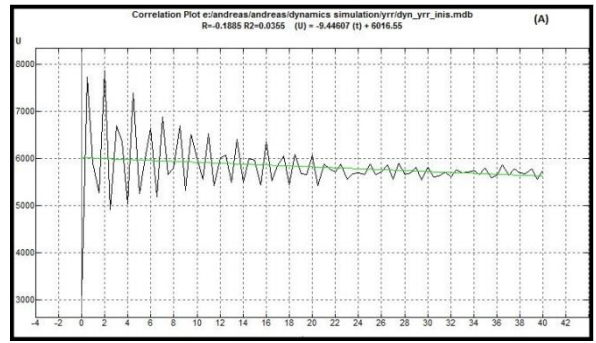

*Fig 5* Initialization curve of E protein-ligand PYRRP complexes

In the protein-ligand complexes, the fluctuation was started from 0 ps to 30 ps for PYRRP ligand (Fig. 5), whereas for ligand PAWRP the fluctuation was started from 0 ps to 28 ps (Fig. 6). The time required to reach the equilibration point were 30 ps for PYRRP protein-ligand complexes and 28 ps for PAWRP protein-ligand complexes. After reaching the point of equilibrium, both protein-ligand complexes were able to adjust the conformation with the implicit solvent.

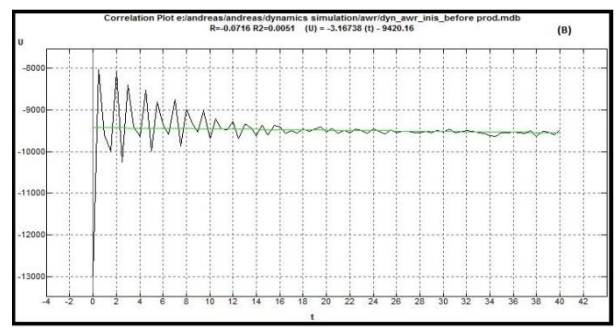

*Fig 6.* Initialization curve of E protein-ligand PAWRPP complexes

The second stage of the production of protein-ligand complexes (PYRRP and PAWRP) was performed at 310 K and 312 K, which represents normal and fever body temperature, respectively. The production stage is divided into three stages, namely heating, the main simulation, and cooling. The heating and cooling stages were conducted to find the lowest energy conformation of the complex.

The result of the production phase at 310 K indicated that the two ligands, PYRRP and PAWRP, retained the hydrogen bond interaction with the envelope cavity until the end of the cooling stage. During the early stage of initialization until the end of 5000 ps simulation stage, the PYRRP and PAWRP ligands still interacted with the binding site residues of envelope protein cavity. Formation of hydrogen bonds between the ligands and the protein cavity during the dynamics simulation occurred in several ways, namely through interaction with the ligand's side chain and the backbone of the protein cavity (Table 6).

The result of the production phase at 312 K also showed that both ligands, PYRRP and PAWRP, also retained interaction with the cavity on the DENV envelope protein until the end of the cooling stage. During the initialization process until the end of 5000 ps simulation stage, the PAWRP and PYRRP binding position were remain consistent with the binding site residues of envelope protein cavity. Hydrogen bonds between the PYRRP ligand and Glu A360, Asp A362, and Ser A363 residues of the envelope cavity remained stable until the end of the simulation. PAWRP also maintained its imteraction with Glu A360 and A147 residuess of the envelope cavity until the end of simulation (Table 7).

*Table 6.* Hydrogen bonds of the complex during molecular dynamics simulation at 310K

| Steps \ Ligand | PYRRP | PAWRP |
| --- | --- | --- |
| Initialization | Glu A360, Glu A360, Glu A360, Glu A360, Asp A362, Asp A362, Asp A362 | Glu A360, Glu A360, Lys A157 |
| Heating 10 ps | Glu A360, Glu A360, Glu A360, Asp A362, Ser A363 | Glu A147, Glu A148, Glu A360 |
| Simulation 5000 ps | His A158, His A158, Lys A295, Lys A295, Asp A329, AspA329, Glu A360, Glu A360, Glu A360, Lys A361, Lys A361, Asp A362, Asp A362, Asp A362, Thr A359, Glu A360, Glu A360, Glu A360, Ser A363 | Glu A148, Glu A148, Glu A148, His A149, Lys A295 |
| Cooling 10 ps | His A158, His A158, Lys A295, Lys A295, Asp A329, Asp A329, Thr A359, Thr A359, Glu A360, Glu A360, Glu A360, Glu A360, Lys A361, Lys A361, Asp A362, Asp A362, Asp A362, Ser A363 | Glu A148, Glu A148, His A149, His A149, Lys A295 |

*Table 7.* Hydrogen bonds of the complex during molecular dynamics simulation at 312K

| Steps \ Ligand | PYRRP | PAWRP |
| --- | --- | --- |
| Initialization | Glu A360, Glu A360, Glu A360, Glu A360, Asp A362, Asp A362, Asp A362 | Glu A360, Glu A360, Lys A157 |
| Heating 10 ps | Glu A360, Glu A360, Glu A360, Asp A362, Ser A363 | Glu A147, Glu A148, Glu A360 |
| Simulation 5000 ps | Glu A147, Glu 360, Glu A360, Glu 360, Ser A363 | Glu A147, Glu 360, Glu A360, Glu 360 |
| Cooling 10 ps | Glu A147, Glu A147, Gly A177, Ser A363, Thr A359, Glu A360, Glu A360, Glu A360 | Glu A147, Glu 360, Glu A360, Glu 360 |

Note : Residues in red are the binding sites of DENV E protein cavity

The magnitude of the protein-ligand complex conformational changes due to the influence of implicit solvent and temperature can be studied from the curve of its simulation time versus RMSD (Root Mean Square Deviation).

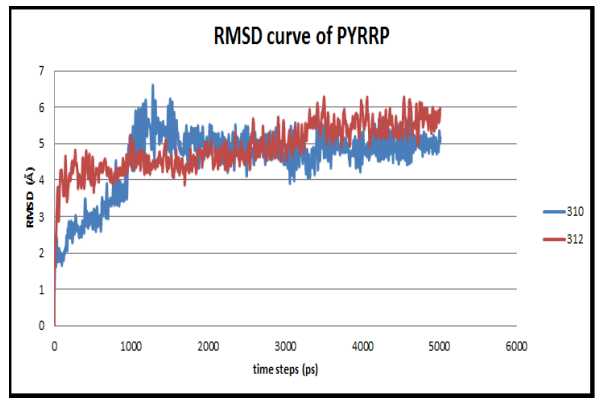

*Fig 7.* The relation between RMSD value of protein-ligand PYRRP complex and time steps during the 5000-ps-molecular–dynamics-simulation

The RMSD curve of protein-PYRRP complex showed that there were fluctuations during 1000-1500 ps simulation at 310 K. However, the complex maintained its stability at 310 K and 312 K in the interval of 2500-5000 ps simulation (Fig. 7). Hence, the PYRRP ligand was able to maintain a stable interaction with the envelope cavity at both tested temperature.

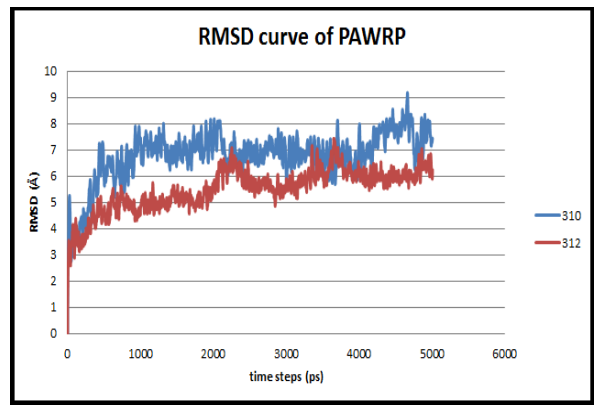

**Fig 8**. The relation between RMSD value of protein-ligand PAWRP complex and time steps during the 5000-ps-molecular–dynamics-simulation

RMSD curve of protein- PAWRP complex at 312 K has a smaller RMSD value than it has at 310 K. This suggests that there are more conformational changes of complex occur at 310 K rather than at 312 K. Therefore, the complex between PAWRP and envelope protein is more stable at 310 than at 312 K.

## 4. CONCLUSION

The screening by means of molecular docking process generated five ligands (PYRRP, PAWRP, PCWRP, PFWRP, and PWPRP) which have better affinity (as indicated by their $\Delta G_{binding}$ and pKi) with envelope cavity than the standards. Based on pharmacological and toxicological prediction analysis, PYRRP and PAWRP were the best ligands among all. During molecular dynamics simulation, both ligands were able to maintain interaction with DENV envelope protein cavity until the end of 5000-ps-simulation, either at 310 K or at 312 K. It was indicated from RMSD analysis that the PYRRP ligand was able to maintain the stability of protein-ligand complex at both tested temperature, while the PAWRP ligand was more reactive at 310 K than at 312 K. Therefore, PYRRP ligand has potential to be developed further as a DENV fusion inhibitor.

**Acknowledgement**

The authors would like to thank Hibah Riset Unggulan UI DRPM/RII/224/RU-UI/2013 for supporting this research. Usman Sumo Friend Tambunan supervised this research, Andreas S Nugroho and Amalia Hapsari worked on the technical details; Arli Aditya Parikesit and Hilyatuz Zahroh prepared the manuscript and re-verified data analysis. The authors would also thank to Dr.Syarifuddin Idrus for the technical assistance.